\documentclass[preprint2]{aastex}

\usepackage{natbib}
\usepackage{graphicx}
\newcommand{\msun}{\mbox{$\,{\rm M}_\odot$}}

\begin{document}

\title{Is the distant globular cluster Pal 14 in a deep-freeze?}

\author{Andreas H.W. K\"upper\altaffilmark{1,2} and Pavel Kroupa\altaffilmark{2}}
\altaffiltext{1}{European Southern Observatory, Alonso de Cordova 3107, Vitacura, Santiago, Chile}
\altaffiltext{2}{Argelander Institut f\"ur Astronomie (AIfA), Auf dem H\"ugel 71, 53121 Bonn, Germany}
\email{akuepper@astro.uni-bonn.de, pavel@astro.uni-bonn.de}

\begin{abstract}
We investigate the velocity dispersion of Pal~14, an outer Milky-Way globular cluster at Galactocentric distance of 71 kpc with a very low stellar density (central density 0.1-$0.2\msun$/pc$^3$). Due to this low stellar density the binary population of Pal~14 is likely to be close to the primordial binary population.\\
Artificial clusters are generated with the observed properties of Pal~14 and the velocity dispersion within these clusters is measured as \citet{Jordi09} have done with 17 observed stars of Pal~14. We discuss the effect of the binary population on these measurements and find that the small velocity dispersion of 0.38 km/s which has been found by \citet{Jordi09} would imply a binary fraction of less than 0.1, even though from the stellar density of Pal~14 we would expect a binary fraction of more than 0.5. We also discuss the effect of mass segregation on the velocity dispersion as possible explanation for this discrepancy, but find that it would increase the velocity dispersion further.\\
Thus, either Pal~14 has a very unusual stellar population and its birth process was significantly different than we see in today's star forming regions, or the binary population is regular and we would have to correct the observed 0.38 km/s for binarity. In this case the true velocity dispersion of Pal~14 would be much smaller than this value and the cluster would have to be considered as ``kinematically frigid'', thereby possibly posing a challenge for Newtonian dynamics but in the opposite sense to MOND.
\end{abstract}

\keywords{binaries: general --- globular clusters: individual (Palomar 14) --- globular clusters: general}

\section{Introduction}
Stellar populations with densities less than about $10^2$ stars/pc$^3$ have a binary fraction,
\begin{equation}
	f_{bin} = \frac{N_{bin}}{N_{bin}+N_{sing}},
\end{equation}
of at least 50 per cent. For example, in the solar neighbourhood (density 1 star/pc$^3$) $f_{bin} \approx 0.6$ \citep{Duquennoy91} or in OB associations (density $< 0.1$ star/pc$^3$) $f_{bin} > 0.7$ \citep{Kouwenhoven07}. In star-forming regions the binary fractions can be even higher, such as in the Taurus-Auriga groups (density about 10 stars/pc$^3$) where the fraction is near 100 per cent (e.g. \citealt{Kroupa03}). But even in high-density star-forming regions, such as the 1 Myr old Orion Nebula cluster which has a density of $10^4$ stars/pc$^3$ \citep{Hillenbrand98}, the binary fraction is near 50 per cent \citep{Koehler06}. Stellar-dynamical models have demonstrated that the Orion Nebula cluster must be expanding and that consequently further destruction of binary systems has mostly ceased \citep{Kroupa01b}. Only in globular clusters, which are dense ($>10^3$ stars/pc$^3$) and which have a long dynamical history over a Hubble time, are binary fractions observed to be less than 10-20 per cent \citep{Hut92}. But this is probably understandable as a result of the destruction of binary systems, first in the very dense initial configuration which these objects typically formed with \citep{Marks08, Marks10}, followed by long-term binary-star burning.\\\\
Pal~14 is an old globular cluster  with a mass of about 10$^4 \msun$, a (three-dimensional) half-light radius of 34 pc, and is located at a distance of 71 kpc from the Galactic centre \citep{Hilker06, Jordi09}. It has a low density (0.1-0.2 stars/pc$^3$) and so conformity with known stellar populations would imply it to have $f_{bin} \approx 0.5$. Measurements of $f_{bin}$ would thus allow testing the dependence of the binary population on the physical conditions during star formation, as Pal~14 is a low-metallicity population that formed nearly a Hubble time ago \citep{Jordi09}. Furthermore, \citet{Jordi09} suggest that Pal~14 may be significantly mass segregated. How mass segregation can affect a velocity dispersion measurement has not been investigated, though.\\\\
However, Pal~14 is also interesting for testing gravitational theories. That binaries can affect the dynamical mass estimate of stellar populations has already been studied in the context of dwarf-spheroidal satellite galaxies by \citet{Hargreaves96} and has recently been re-addressed by \citet{Kouwenhoven08} and \citet{Gieles10} for star clusters in the Milky Way disk. The measured (one-dimensional) velocity dispersion of Pal~14 is ($0.64\pm0.15$) km/s or ($0.38\pm0.12$) km/s, depending on whether the outlying ``star 15'' of a sample investigated by \citet{Jordi09} is included or not, respectively. Now, if Pal~14 has a binary fraction of 50 per cent or more, then the velocity dispersion measured from a single spectroscopic snapshot may be significantly contaminated, i.e. inflated, because of the binary-star orbital motions of the tracer stars. A potential problem arising here is that the dynamical velocity dispersion of the cluster must be about 0.5 km/s for Newtonian dynamics to hold. So there is not much room for a binary-star contribution to the measured velocity dispersion. Alternatively, if the binary population is normal (i.e. canonical), then the true (binary-corrected) velocity dispersion of the virialised cluster may be much smaller than the Newtonian value. If this were the case, then a Newtonian crisis would emerge, such that Pal~14 would be ``kinematically frigid'', a situation which is not expected from current theoretical dynamics. A third possibility is given by the unreliability of low-number statistics in such systems, i.e. a sample of about 17 stars may simply be insufficient to determine the velocity dispersion of such a stellar system.\\\\
Thus, either Pal~14 is in virial equilibrium, such that its stellar population would be non-canonical (a binary fraction much lower than all other known populations). Or, the cluster has a canonical stellar population, but is then kinematically frigid, in violation of Newtonian dynamics. This would not be remedied by MOND, because according to MOND the cluster ought to have an even larger velocity dispersion than the Newtonian value \citep{Baumgardt05}. The third explanation, of low-number statistics, would imply that velocity dispersion measurements of low-velocity stellar systems would have to be taken with much more caution than is done in practice nowadays.\\\\
With this contribution we assess the expected velocity dispersion of Pal~14 assuming various binary fractions and Newtonian dynamics to be valid (Section 2). These numerical experiments are compared to the observed velocities in the cluster in Section 3. The conclusions are contained in Section 4. 

\section{Velocity dispersions}\label{Sec:Vel}
\begin{figure}
\plotone{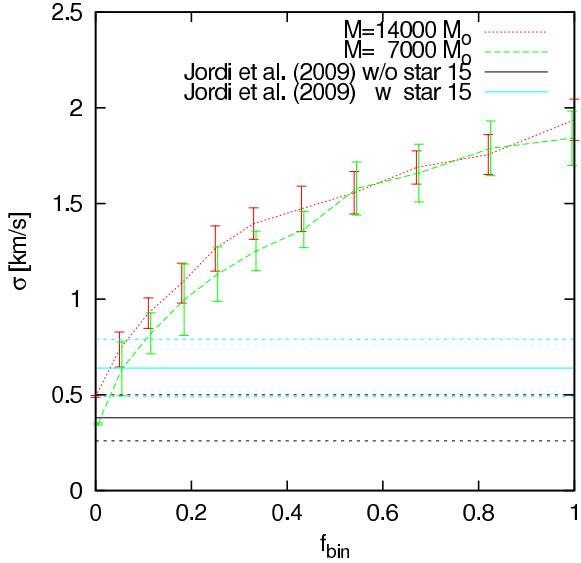}
  \caption{Mean velocity dispersions of the cluster renditions with $7000\msun$ and $14000\msun$ as determined from all stars above $0.7\msun$ for ten different binary fractions, $f_{bin}$. The error bars give the standard deviation from the mean. The velocity dispersion can increase by a factor of 5 for large binary fractions, compared to the binary free case. Above $f_{bin} = 0.5$ the binaries dominate the velocity dispersion such that the two cluster groups get indistinguishable. Also shown are the measured results from \citet{Jordi09} with and without taking into account star 15 (see text). Their velocity dispersion measurements allow a binary fraction of less than 0.1, or 0.2 taking into account star 15, respectively.}
  \label{sigma_all}
\end{figure}

\begin{figure}
\plotone{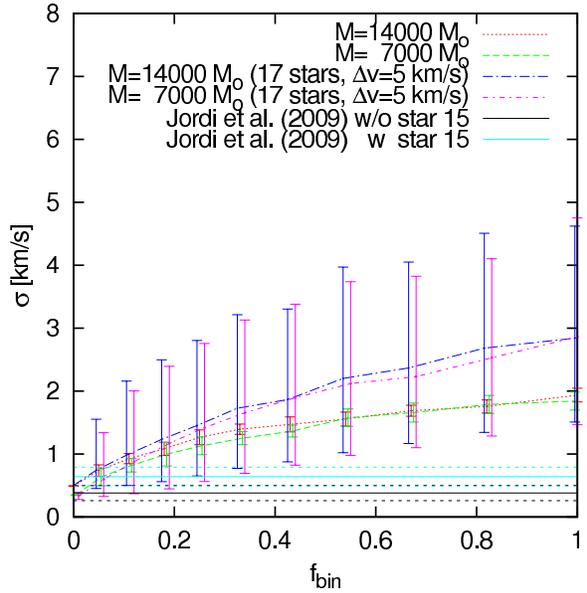}
  \caption{Same as Fig.~\ref{sigma_all} but additionally the velocity dispersions are shown which are derived from sub-sets of 17 stars above $0.7\msun$. For this measurement all stars lying outside a velocity interval about the mean value of $\Delta v = 5$ km/s were rejected. The data points show the mean of $3\,000\,000$ renditions, and the error bars show the values in which 68 per cent of all renditions lie. The results of \citet{Jordi09} allow a binary fraction of less than about 0.1, or 0.3 taking into account star 15, respectively.}
  \label{sigma_5kms}
\end{figure}

\begin{figure}
\plotone{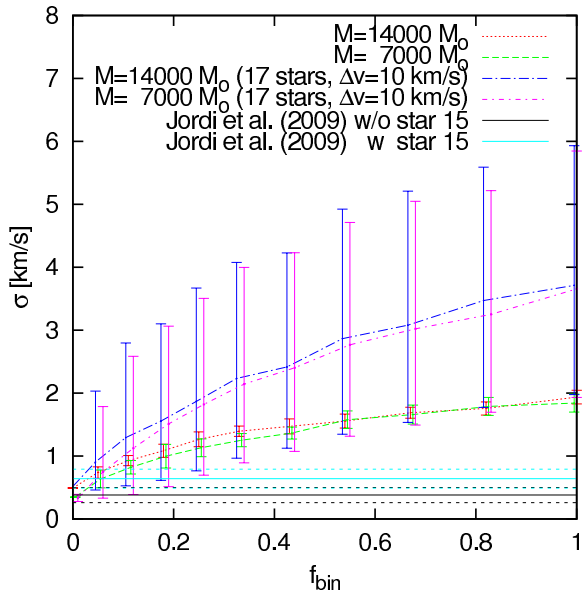}
  \caption{Same as Fig.~\ref{sigma_5kms} but for a velocity interval of $\Delta v = 10$ km/s. The effects from Fig.~\ref{sigma_5kms} get even more significant. The mean derived velocity dispersion from our artificial clusters can be as large as 10 times the Newtonian velocity dispersion in the case without binaries, and is always significantly larger than the true velocity dispersion within the cluster. Even though the spread about this mean value is large, at $f_{bin} = 1.0$ the velocity dispersion of the sub-sample and of all stars barely agree within the error bars.}
  \label{sigma_10kms}
\end{figure}

\begin{figure}
\plotone{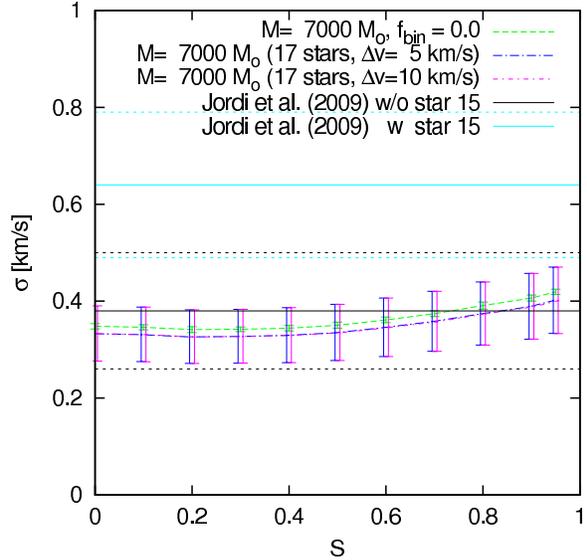}
  \caption{Mean velocity dispersions of the cluster renditions with $7000\msun$ as determined from all stars above $0.7\msun$ (as in Fig.~\ref{sigma_all}), as well as derived from sub-sets of 17 stars above $0.7\msun$ in a velocity intervall of  $\Delta v = 5$km/s (Fig.~\ref{sigma_5kms}) and  $\Delta v = 10$ km/s (Fig.~\ref{sigma_10kms}), respectively, but here for ten different degrees of mass segregation, $S$, as defined by \citet{Subr08}. For increasing mass segregation the velocity dispersion rises as heavy stars are preferentially located in the cluster centre and hence move at a higher velocity on average, thus mass segregation cannot explain the low velocity dispersion measurement of \citet{Jordi09}.}
  \label{sigma_ms}
\end{figure}

For the purpose of testing the observed velocity dispersion of Pal~14, we create artificial star clusters with the observed properties of Pal~14 and measure the velocity dispersion in these clusters. Therefore we use the code \textsc{McLuster}\footnote{\tt www.astro.uni-bonn.de/\~{}akuepper/mcluster/mcluster.html} (K\"upper et al., in prep.), an open-source project developed at AIfA Bonn to set up clusters for $N$-body computations, especially for \textsc{Nbody6} \citep{Aarseth03}. We use this software to conveniently generate 300 renditions of Pal~14 with various properties for taking out mock observations and testing the findings of \citet{Jordi09} on their conclusiveness.\\\\
The parameters for the artificial clusters we take from \citet{Jordi09} and \citet{Hilker06}. The most important quantity in this respect is the total mass of the cluster as it should be closely linked to the velocity dispersion in the assumed case of the cluster being in virial equilibrium. Using star counts, \citet{Jordi09} find a mass for Pal~14 between $6000-12000\msun$ without taking into account compact remnants, and depending on the assumed mass function within the cluster. Assuming additional 15-20 per cent mass in stellar remnants (\citealt{Dabringhausen09}, and Dabringhausen private communication) yields a mass range for Pal~14 of $7000-14000\msun$. We therefore concentrate on the two cases of $7000 \msun$ and $14000\msun$, respectively.\\\\ 
As the density distribution we choose a Plummer sphere with a half-mass radius of 32 pc, which is the observed half-light radius of Pal~14 of 1\farcm28 \citep{Hilker06} at the assumed distance of ($71\pm 1.3$) kpc \citep{Jordi09}. Setting the half-light radius equal to the half-mass radius assumes that mass follows light, just as \citet{Jordi09} have done in their investigation, i.e. that the cluster is not significantly mass segregated. Moreover, the infinite Plummer distribution is cut off at the Jacobi radius ($\sim 128/156$ pc for $7000/14000\msun$) assuming a Galactic circular velocity of 220 km/s at the Galactocentric radius of Pal~14.\\\\
As the mass function of the cluster stars we use the canonical IMF \citep{Kroupa01a}, but cut it off at a maximum stellar mass of $1.0\msun$ as all stars above this mass should have died by stellar evolution at the expected age of Pal~14 of 11.5 Gyr \citep{Jordi09}. Even though \citet{Jordi09} find a shallower slope in the range of $0.5\msun$ to $0.8\msun$, we argue that the actual slope of the mass function is not very important in this respect as the velocities of the cluster stars get drawn independently of each other and we, in the end, only observe stars of mass above $0.7\msun$ to be consistent with \citet{Jordi09}.\\\\
For investigating the effect of a realistic binary population on the velocity dispersion, we set up the binary population following the \citet{Duquennoy91} period distribution for field stars and a thermal eccentricity distribution. The binary orbital planes are distributed randomly, as are the orbital phases. We reject binaries with a semi-major axis below 100 $R_\sun$, though, as these binaries may have been destroyed in a common-envelope phase of the binary components \citep{Zorotovic}. Moreover, we use random pairing for the binary components since we are dealing with an evolved population of low-mass stars. We vary the binary fraction, $f_{bin}$, from 0 to 1 and generate ten clusters for each binary fraction to gain better statistics.\\\\
In addition, we set up clusters with the above properties but being mass segregated. Therefore, we use the procedure defined in \citet{Subr08} which is implemented through \textsc{Plumix} in \textsc{McLuster}. \citet{Subr08} defines the degree of mass segregation through a single parameter, $S$, which can vary from 0 (not segregated) to 1 (completely mass segregated). To keep the number of models to a minimum, we concentrate on the cluster with $7000\msun$ without binaries, and see how mass segregation affects its velocity dispersion. Again we generate 10 clusters for each value of $S$.\\\\
The line-of-sight velocity dispersion, $\sigma$, we measure from our clusters by calculating
\begin{equation}\label{eq:sigma}
\sigma = \sqrt{\overline{v^2}-\overline{v}^2},
\end{equation}
where $\overline{v}$ is the mean velocity in the sample, and $\overline{v^2}$ is the mean squared velocity. We do this separately in three different directions and take each as an independent measurement.\\\\
In Fig.~\ref{sigma_all} we show the mean velocity dispersions for all stars above $0.7\msun$ for ten different binary fractions and for both cluster masses. The error bars give the standard deviation of the different cluster renditions from the mean. The clusters without binaries have a very little spread about the mean whereas already a binary fraction of 0.1 introduces such a large scatter that the two mass groups of clusters overlap within their error bars. At a binary fraction of 0.5 the two mass groups are indistinguishable as their velocity dispersion gets completely dominated by the binary population. At a binary fraction of 1.0 the velocity dispersion is about four times larger than in the case without binaries.\\\\
In Fig.~\ref{sigma_5kms} we show the velocity dispersion when we draw 17 stars out of the total sample of stars above $0.7\msun$, just as \citet{Jordi09} have done. The data points show the mean of $3\,000\,000$ renditions out of the ten clusters for each binary fraction. Again we took the line-of-sight velocity dispersions independently along three different directions for each rendition. Since the final distribution of velocity dispersions for each binary fraction does not follow a Gaussian distribution but is rather asymmetric, the error bars show 68 per cent of all renditions below the mean and 68 per cent of all renditions above the mean.\\\\
We rejected stars, though, when their velocities were off the mean by more than 2.5 km/s, i.e. only took stars within a velocity window of $\Delta v = 5$ km/s into account. This is similar to the way \citet{Jordi09} get their velocity dispersion of Pal~14, and is the usual practise to discriminate between cluster members and non-members. We see that the mean measured velocity dispersion can increase by about 50 per cent compared to the velocity dispersion of all stars above $0.7\msun$ if we only have 17 stars in our sample. This effect is small for low binary fractions but gets significant above a binary fraction of about 0.3. Moreover, the scatter in our velocity dispersion measurement is huge. At $f_{bin} = 0.5$ we can measure values between 1 km/s and 3 km/s, depending on which stars we take into our sample. At a binary fraction of 1.0 the mean measured velocity dispersion is about 7 times larger than the Newtonian velocity dispersion without binaries.\\\\
In Fig.~\ref{sigma_10kms} we show the same experiment but with a velocity window of $\Delta v = 10$ km/s. The effect described above gets more significant. The mean value of the measured velocity dispersion increases and also the scatter grows. At $f_{bin} = 0.5$ we can get values between 1 km/s and 4 km/s. At a binary fraction of 1.0 the measurements of a small sub-sample and the measurements of all stars barely agree within the error bars any more. Note that the mean value can be as large as 10 times the true Newtonian velocity dispersion without binaries. Moreover, from comparing Fig.~\ref{sigma_5kms} to Fig.~\ref{sigma_10kms} we see that the measured velocity dispersion depends strongly on the velocity window which we allow. We tend to increasingly overestimate the true velocity dispersion by increasing the size of the allowed velocity window.\\\\
In Fig.~\ref{sigma_ms} the velocity dispersion measurements are shown for the mass segregated clusters. The plot shows the velocity dispersion measurements for a cluster of $7000\msun$ without binaries, as seen in Fig.~\ref{sigma_all}-\ref{sigma_10kms} but for a mass segregation degree, $S$, varying from 0 to 0.95. From the figure we see that mass segregation increases the velocity dispersion further. The segregated cluster ($S=0.95$) has a 20 per cent higher velocity dispersion than the unsegregated case ($S=0$). This is due to the fact that through mass segregation the heaviest stars (which we observe for the velocity dispersion) move to the cluster centre, and that stars on average move at higher velocities when they are in the cluster centre. When a sample of observed stars was concentrated on the centre of the cluster and was not well distributed over the cluster, we would get the same effect.

\section{Discussion}\label{Sec:Dis}
\citet{Jordi09} have determined the velocity dispersion of Pal~14 according to \citet{Pryor93}, i.e. have made a maximum-likelihood estimation, which is, of course, necessary as unlike in our samples the measurement errors of their radial velocities are all different. From their sample of 17 stars they find one star (star 15) to lie 3 $\sigma$ off the mean value, and thus split their investigation into two parts: one with taking star 15 into account and one without taking star 15 into account. For the sample with star 15 they find a velocity dispersion of (0.64$\pm$0.15) km/s, and without star 15 they find (0.38$\pm$0.12) km/s.\\\\
Based on a Kolmogorov-Smirnov test, \citet{Jordi09} argue that star 15 is most likely not a regular member of Pal~14, i.e. a foreground contamination or part of a long-period binary system. Thus, at the bottom line they favour the lower value of 0.38 km/s.\\\\
From Fig.~\ref{sigma_all} we see that such a low velocity dispersion would imply that the binary fraction of Pal~14 was less than 0.1. With taking star 15 into account the velocity dispersion of Pal~14 would imply a binary fraction of less than 0.2.\\\\
In Fig.~\ref{sigma_5kms} the velocity dispersion derived from a sub-sample of 17 stars within a velocity interval of $\Delta v = 5$ km/s shows that the lower value of 0.38 km/s would be consistent within the error bars with values less than $f_{bin} = 0.2$, while the higher value of 0.64 km/s would be consistent with values less than $f_{bin} = 0.4$. Allowing for $\Delta v = 10$ km/s reduces the latter to $f_{bin}$ less than 0.3 (Fig.~\ref{sigma_10kms}).\\\\
From Fig.~\ref{sigma_ms} we see that mass segregation does not help in understanding the low observational value of Pal~14's velocity dispersion, since mass segregation tends to increse the observed velocity dispersion of a cluster by up to 20 per cent compared to the unsegregated case. On the contrary, since Pal~14 is supposed to be mass segregated \citep{Jordi09}, its velocity dispersion may be even inflated. Our previous estimates on Pal~14's binary fraction therefore have to be taken as upper limits.\\\\

\section{Conclusions}\label{Sec:Conclusions}
From our test of the observed velocity dispersion of the Milky-Way globular cluster Pal~14, we have seen that the binary fraction within Pal~14 has to be less than 0.2 in the case of star 15 not being considered a member, in order to be consistent with a velocity dispersion as low as (0.38$\pm$0.12) km/s. With taking star 15 into account the maximum binary fraction consistent with the observational uncertainties is about 0.4. This poses a number of questions on the nature of Pal~14.\\\\
As a first explanation of these findings we may assume that the binary fraction within Pal~14 is indeed as low as found. But, as stated above, the density within Pal~14 is as low as 0.1-0.2 stars/pc$^3$. The effect of disruption of binaries due to dynamical stellar evolution within its age of about 11.5 Gyr is therefore negligible and we even may see here the primordial binary population of a globular cluster \citep{Hasani10}. Thus, the formation of Pal~14 must have been significantly different from what is observed in all other star forming sites today. Or Pal~14 must have undergone a very dense and violent phase in which most of the binaries were burned, but as we can see in the ONC today, this is very unlikely to happen for a sufficiently long time span that the binary fraction drops below 0.5. Furthermore, such a scenario would give rise to the question how Pal~14 could have expanded that much, as with a half-light radius of about 34 pc it is one of the most expanded globular clusters of the Milky Way today. Recent numerical studies moreover show that this expansion is very unlikely to be of pure dynamical origin, since expansion takes place on a relaxation time scale and the relaxation time of Pal~14 is of order of a Hubble time \citep{Hasani10}.\\\\  
A second option would be that the binary population is normal, i.e. above 0.5, and that thus the observed velocity dispersion has to be corrected for the effect of binaries. As we have seen, this can be as much as a factor of 10 in the case of 17 stars drawn from a cluster with a binary fraction of 1.0. Since the lower mass limit of Pal~14 is determined to be about $7000\msun$ \citep{Jordi09} this would imply that the true velocity dispersion of Pal~14 is much lower than the Newtonian prediction, i.e. Pal~14 is ``kinematically frigid''. This would be inconsistent with our understanding of Newtonian gravity and could neither be explained by considering MOND to be valid in Pal~14.\\\\
Moreover, we found that mass segregation increases the observed velocity dispersion of a cluster even further, and thus cannot explain the low velocity dispersion of Pal~14. The observed, flattened mass function of Pal~14 on the other hand suggests that Pal~14 is significantly mass segregated \citep{Jordi09}. Thus, its unsegregated, i.e. for the effect of mass segregation corrected, velocity dispersion may be even lower than the values reported by \citet{Jordi09}, which would worsen the problem.\\\\ 
Thus, the present state of knowledge on Pal~14 is that either its binary fraction is highly abnormal, given its low density, or that it is significantly sub-virial. The former possibility would imply a non-standard star-formation event which formed Pal~14, while the latter indicates a problem understanding the dynamics of Pal~14.\\\\
In any case, this investigation has shown that one has to be cautious with low velocity dispersions derived from small samples of cluster stars. The observed value from such a sub-sample tends to be significantly larger than the true velocity dispersion. This effect gets larger with increasing size of the allowed velocity window about the mean radial velocity. This could be especially important in the outer parts of star clusters (e.g. \citealt{Scarpa07}), as was also recently shown by \citet{Kuepper10}.\\\\
More line-of-sight velocity measurements, e.g. as a 2nd epoch spectroscopic snapshot, may help to reduce the statistical uncertainties and improve the significance of findings on whether or not Newtonian dynamics is valid in Pal~14 (see also \citealt{Gentile10}).

\acknowledgements
The authors would like to thank Thijs Kouwenhoven, who helped to significantly increase the scope of this work. AHWK kindly acknowledges the support of an ESO Studentship.

\clearpage

\clearpage

\end{document}